\documentstyle{article}
\begin{document}
\rightline{physics/9707012}
\begin{center}
{\bf Supersymmetric partner chirping of Newtonian free damping}\\

H. Rosu$^{a}$\footnote{E-mail: rosu@ifug3.ugto.mx},
%
J. L. Romero$^{b}$ and J. Socorro$^{a}$\footnote
{E-mail: socorro@ifug4.ugto.mx}

$^{a}$
Instituto de F\'{\i}sica de la Universidad de Guanajuato,
Apdo Postal E-143, 37150 Le\'on, Gto, M\'exico\\

$^{b}$
Departamento de F\'{\i}sica de la Universidad de Guadalajara,
Corregidora 500, 44100 Guadalajara, Jal, M\'exico
\end{center}



\bigskip

{\bf Summary}. - We connect the classical free damping cases by means of
Rosner's construction in supersymmetric quantum mechanics.
Starting with the critical damping, one can obtain in the
underdamping case a chirping of
instantaneous physical frequency $\omega ^{2}(t)
\propto \omega _{u}^{2}{\rm sech}^2(\omega _{u}t)$, whereas in the
overdamped case the ``chirping" is of the (unphysical)
type $\omega ^{2}(t)\propto\omega _{o}^{2}{\rm sec}^{2}(\omega _{o}t)$,
where $\omega _{u}$ and $\omega _{o}$ are the underdamped and overdamped
frequency parameters, respectively.

\bigskip
\bigskip



\bigskip





\bigskip
\bigskip

The techniques of supersymmetric quantum mechanics gave a new impetus to
many research areas in the last two decades \cite{1}.
In this note, we start with the critical free damping case in classical
mechanics and construct the corresponding supersymmetric partners.
The free damping Newton equation reads
$$
m \frac{d^2x}{dt^2}+\gamma \frac{dx}{dt}+k x=0~.
\eqno(1)
$$
Using the gauge change of dependent
variable $x=y\exp (-\frac{\gamma}{2m}t)$,
this equation can be put in the following Schr\"odinger form in the time domain
$$
y^{''}-\Big[\Big(\frac{\gamma}{2m}\Big)^2-\frac{k}{m}\Big]y=0~,
\eqno(2)
$$
or $y^{''}-\omega _{d}^{2}y=0$, where $\omega _{d}^{2}=(\gamma/2m)^2-k/m$.
Thus, one can discuss separately, the classical
cases of underdamping (oscillating relaxation),
critical damping (rapid nonoscillating relaxation),
and overdamping (slow  nonoscillating relaxation), i.e.,
$-\omega _{u}^{2}=\omega _{d}^{2}<0$,
$\omega _{c}^{2}=\omega _{d}^{2}=0$ and $\omega _{o}^{2}=\omega _{d}^{2}>0$,
respectively \cite{mec}. Notice that the physical $\omega _{d}^2$ frequencies
are negative, the positive ones are only convenient mathematical symbols
corresponding in fact to nonoscillating regimes.

We now proceed with the supersymmetric scheme that we apply in a manner
similar to Rosner \cite{r}. Thus, we start with the case of critical damping
corresponding in quantum mechanics to a potential which is zero and relate it
to a Schr\"odinger-type equation in the time domain which has a
fundamental frequency at $-\omega _{u}^{2}$,
i.e., a single oscillating relaxation mode that
we consider as the equivalent of a bound state in the usual
quantum mechanics. In other words, we solve the ``fermionic" Riccati equation
$$
W^2_{1}-W^{'}_{1}+\omega _{d}^{2}=0~,
\eqno(3)
$$
i.e.,
$$
W^2_{1}-W^{'}_{1}-\omega _{u}^{2}=0~,
\eqno(4)
$$
to find Witten's superpotential $W_1(t)=-\omega _{u}\tanh [\omega _{u}t]$
and next go to the ``bosonic" Riccati equation
$$
W^2_{1}+W^{'}_{1}+\omega _{1}^{2}(t)-\omega _{u}^{2}=0~,
\eqno(5)
$$
in order to get $\omega _{1}^{2}(t)=-2\omega _{u}^{2}
{\rm sech}^2[\omega _{u}t]$. Moreover, one can write the Schr\"odinger
equation corresponding to the ``bosonic" Riccati equation as follows
$$
-\tilde{y}^{''}+\omega _{1}^{2}(t)\tilde{y}=-\omega _{u}^{2}\tilde{y}~,
\eqno(6)
$$
with the localized solution
$\tilde{y}\propto \omega _{u}{\rm sech}(\omega _{u}t)$.
The physical picture is that of a chirping ${\rm sech}$ soliton profile
containing a
single oscillating relaxation mode self-trapped at $-\omega _{u}^2$
within the frequency pulse. One can
employ the scheme recursively to get
several oscillating relaxation modes embedded in the chirping frequency
profile.  
Indeed, suppose we would like to introduce $N$ oscillating relaxation
modes of the type $\omega _{n}^{2}=-n^2\omega _{u}^{2}$, $n=1,...N$ in the
${\rm sech}$ chirp. Then, one has to solve the sequence of equations
$$
W_{n}^{2}-W_{n}^{'}=\omega _{n-1}^2+n^{2}\omega _{u}^{2}
\eqno(7a)
$$
$$
W_{n}^{2}+W_{n}^{'}=\omega _{n}^2+n^{2}\omega _{u}^{2}
\eqno(7b)
$$
inductively for $n=1...N$ \cite{r}. The chirp frequency
containing $N$ underdamped
frequencies $n^2\omega _{u}^{2}$, $n=1...N$ is of the form
$\omega _{N}^{2}(t)=-N(N+1)\omega _{u}^{2}{\rm sech}^2(\omega _{u}t)$. The
relaxation modes can be written in a compact form as follows
$$
\tilde{y}_{n}(t;N)\approx A^{\dagger}(t;N)A^{\dagger}(t;N-1)
A^{\dagger}(t;N-2)...A^{\dagger}(t;N-n+2){\rm sech} ^{N-n+1}\omega _{u}t~,
\eqno(8)
$$
i.e., by applying the first-order operators
$A^{\dagger}(t;a_{n})=-\frac{d}{dt}-a_{n}\omega _{u}\tanh (\omega _{u}t)$,
where $a_{n}=N-n$, onto the ``ground state" underdamped mode.
This scheme can be easily generalized to embedding frequencies of the
type $-\omega _{u,i}^2=(\gamma _{i}/2m)^2-k/m$ and moreover, to the
construction of chirp profiles having a given continuous spectrum of
relaxational modes but we shall not pursue this task here.

On the other hand, in the case of overdamping the ``fermionic"
Riccati equation
$$
W^2_{1}-W^{'}_{1}+\omega _{o}^{2}=0
\eqno(9)
$$
leads to the solution $W_{1}=\omega _{o}{\rm tan}(\omega _{o}t)$ and from
the ``bosonic" Riccati equation
$$
W^2_{1}+W^{'}_{1}+\omega _{1}^{2}+\omega _{o}^{2}=0~,
\eqno(10)
$$
one will find $\omega _{1}^{2}(t)
=2\omega _{o}^{2}{\rm sec}^{2}(\omega _{o}t)$. Consequently, the Schr\"odinger
equation
$$
-\tilde{y}^{''}+\omega _{1}^{2}(t)\tilde{y}=\omega _{o}^{2}\tilde{y}
\eqno(11)
$$
has solutions of the type
$\tilde{y}\propto \omega _{o}{\rm sec}(\omega _{o}t)$, and therefore the
approach leads to unphysical results.

Referring again to the underdamped case, we also
remark that an interesting, polar analysis of the chirp frequency profile can
be performed by means of the change of variable
$t={\rm ln}({\rm tan}\frac{\theta}{2})$ leading to an associated Legendre
equation in the spherical polar coordinate $\theta$
$$
\frac{d^2\tilde{y}}{d\theta ^2}+\cot \theta\frac{d\tilde{y}}{d\theta}+
\Big[N(N+1)-\frac{n^{2}}{\sin ^{2}\theta}\Big]\tilde{y}=0~.
\eqno(12)
$$

Finally, we mention that there may be potential applications of
supersymmetric approaches to
chirping phenomena in many areas, such as
semiconductor laser physics \cite{book},
the propagation of chirped optical  solitons in fibers \cite{Malo},
and optimal control of quantum systems by chirped pulses \cite{Ams}.

\bigskip

{\bf Acknowledgment}

The work was supported in part by
the CONACyT Projects No. 4868-E9406, No. 3898-E9608,
and a ``Scientific Summer" grant from the University of Guanajuato.


\end{document}